\documentclass[twocolumn,twoside,pra,superscriptaddress]{revtex4}
\usepackage[utf8]{inputenc}

\usepackage[pdftex]{graphicx,color}
\usepackage{amsmath}
\usepackage{amssymb}
\usepackage{epsfig}
\usepackage{amsthm}
\usepackage{mathtools}
\usepackage{tikz}
\usepackage{wrapfig}
\usepackage{comment}
\usepackage{simpler-wick}
\usepackage{simplewick}
\usepackage{latexsym,revsymb}
\usepackage[unicode=true,pdfusetitle,
bookmarks=true,bookmarksnumbered=false,bookmarksopen=false,
breaklinks=false,pdfborder={0 0 0},backref=false,colorlinks=false]
{hyperref}

\DeclareMathOperator*{\argmax}{argmax}

\newcommand{\ket}[1]{| #1 \rangle}

\newcommand{\vertiii}[1]{{\left\vert\kern-0.25ex\left\vert\kern-0.25ex\left\vert #1 
    \right\vert\kern-0.25ex\right\vert\kern-0.25ex\right\vert}}

\makeatletter
\newtheorem*{rep@theorem}{\rep@title}
\newcommand{\newreptheorem}[2]{%
\newenvironment{rep#1}[1]{%
 \def\rep@title{#2 \ref{##1}}%
 \begin{rep@theorem}}%
 {\end{rep@theorem}}}
\makeatother

\newtheorem{theorem}{Theorem}
\newreptheorem{theorem}{Theorem}

\newreptheorem{corollary}{Corollary}

\newreptheorem{lemma}{Lemma}

\begin{document}
\title{Parallel decoding of multiple logical qubits in tensor-network codes}
\author{Terry Farrelly}
\email{farreltc@tcd.ie}
\affiliation{ARC Centre for Engineered Quantum Systems, School of Mathematics and Physics, The University of Queensland, St Lucia, QLD, 4072, Australia}
\author{Robert J. Harris}
\affiliation{ARC Centre for Engineered Quantum Systems, School of Mathematics and Physics, The University of Queensland, St Lucia, QLD, 4072, Australia}
\author{Nathan A. McMahon}
\affiliation{ARC Centre for Engineered Quantum Systems, School of Mathematics and Physics, The University of Queensland, St Lucia, QLD, 4072, Australia}
\affiliation{Friedrich-Alexander University Erlangen-Nürnberg (FAU), Department of Physics, Erlangen, Germany}
\author{Thomas M. Stace}
\affiliation{ARC Centre for Engineered Quantum Systems, School of Mathematics and Physics, The University of Queensland, St Lucia, QLD, 4072, Australia}
\begin{abstract}
We consider tensor-network stabilizer codes and show that their tensor-network decoder has the property that independent logical qubits can be decoded in parallel.  
As long as the error rate is below threshold, we show that this parallel decoder is essentially optimal.  
	As an application, we verify this for the max-rate holographic Steane (heptagon) code.  
	For holographic codes this tensor-network decoder was shown to be efficient with complexity polynomial in $n$, the number of physical qubits.  
	Here we show that, by using the parallel decoding scheme, the complexity is also linear in $k$, the number of logical qubits.
	Because the tensor-network contraction is computationally efficient, this allows us to exactly contract tensor networks corresponding to codes with up to half a million qubits.
	Finally, we calculate the bulk threshold (the threshold for logical qubits a fixed distance from the code centre) under depolarizing noise for the max-rate holographic Steane code to be $9.4$\%.  
\end{abstract}

\maketitle

\section{Introduction}
A significant problem in quantum error correction is the problem of decoding, i.e., finding the best correction operation, given the results of the syndrome measurements.  Indeed, this is known to be computationally difficult ({\#}P-complete for stabilizer codes) \cite{HG11,IP13}.  
(Here we are considering decoding with perfect syndrome measurements.)  While there are many different approaches to decoding, one possibility is via tensor networks \cite{BC17}.  A prominant example 
involved matrix-product states to approximately calculate the probabilities for the maximum likelihood decoder---the optimal decoder given the syndrome---for surface codes \cite{BSV14,TDC19,TBF18,CF18,HHK20}.  
Another method using tensor networks for the surface code used projected entangled pair states to represent the state of the code and subsequently find the resulting error channel on the logical degree of freedom \cite{DP17,DP18}.  Another tensor-network decoder used tensor networks to represent encoding unitaries \cite{FP14,FP14a}.

An exciting, though so far unrelated, development in quantum error correction using tensor networks are holographic codes, which drew inspiration from the AdS/CFT correspondence \cite{FYH15,LS15,HNQ16,Evenbly17,JGP19,JGP19a,KC19,OS20,JZE20}.  In that case, there is a geometric interpretation of the code:\ bulk degrees of freedom in a two-dimensional hyperbolic space correspond to logical qubits, and boundary degrees of freedom correspond to physical qubits.  These codes form toy models of the AdS/CFT correspondence, which comprise strongly coupled gravity in the bulk of AdS spacetime, and an equivalent conformal field theory on the boundary of the spacetime \cite{Harlow16}.  The holographic codes introduced in \cite{FYH15} were built out of so-called perfect tensors.  This restriction to perfect tensors was later weakened to include block-perfect tensors (or perfect tangles) \cite{HMBS18,BO18}.  An example of block-perfect holographic codes includes the Steane holographic code (also known as the heptagon code) \cite{HMBS18}, which uses the seven-qubit Steane code \cite{Steane96} as a building block.  In \cite{HC20}, this code was decoded using an integer optimization method finding a threshold of around $7$\% under depolarizing noise.

In this work, we consider tensor-network stabilizer codes \cite{FHM20}, which naturally come with a tensor-network decoder.  Here we find that this decoder parallelizes:\ to decode some logical qubits in a code optimally (minimizing word error probability), each logical qubit can be decoded separately as long as the physical error rate is below threshold.  This is only useful for bulk logical qubits, by which we mean logical qubits a fixed distance from the code centre.  In contrast, for a logical qubit at the code boundary, there will be no threshold, as the support of its logical operators is fixed.  Note that we are not considering syndrome errors, so this is not a fully fault-tolerant scheme.  
Nevertheless, this allows us to find the best possible performance of the code with respect to any decoder because the maximum likelihood decoder (implemented here using tensor networks) is optimal.

One application of tensor-network codes is to holographic codes.  So, as an example, we apply these methods to calculate the threshold for the central logical qubit in the Steane holographic code, getting a value of $9.4\%$ under depolarizing error (somewhat higher than the result of $7$\% from \cite{HC20}, which used a different decoder).  
We also see that (due to the homogeneity of the code) the threshold for bulk logical qubits, i.e., those a fixed distance from the code centre, has the same value.  
Furthermore, a consequence for holographic codes is that the complexity of decoding any $K\leq k$ logical qubits is linear in $K$ and polynomial in $n$, the number of physical qubits, as the tensor-network contractions used in decoding have polynomial compexity \cite{FHM20}.  Since each logical qubit can be decoded in parallel, the total decoder runtime is independent of $K$.

\section{Stabilizer codes}
\label{sec:Stabilizer}
Let $\sigma^{0}=\openone$, $\sigma^{1}=X$, $\sigma^{2}=Y$ and $\sigma^{3}=Z$ be the the identity operator and the Pauli operators on a single qubit respectively.
Denote the $n$-qubit Pauli group by $\mathcal{G}_n$, which comprises all operators of the form $z\sigma^{i_1}\otimes...\otimes\sigma^{i_n}$, with $z\in\{\pm 1,\pm i\}$.  Stabilizer codes \cite{Gottesman97,NielsenChuang,Roffe19} have the property that logical operators and stabilizers are elements of $\mathcal{G}_n$.  
Logical information is encoded in a subspace of the $n$-qubit Hilbert space that is fixed by the stabilizers---an abelian group $\mathcal{S}\subset\mathcal{G}_n$.  So for any state $\ket{\psi}$ in the code space, $S\ket{\psi}=\ket{\psi}$ for each $S\in\mathcal{S}$.
The stabilizer group has $n-k$ independent generators denoted $S_i$, and the logical subspace has dimension $2^k$, corresponding to $k$ logical qubits.

Logical operators compise a non-abelian group $\mathcal{L}\subset\mathcal{G}_n$.  This group can be generated by $k$ $X$-type and the $k$ $Z$-type operators, denoted by $X_{\alpha}$ and $Z_{\alpha}$, with $\alpha\in\{1,..,k\}$.  All logical operators commute with stabilizers, while satisfying $X_{\alpha}Z_{\beta}=(-1)^{\delta_{\alpha \beta}}Z_{\alpha}X_{\beta}$.

Although the decoder and the results in all later sections apply to general independent Pauli noise (and even locally correlated noise), in the examples we will only consider i.i.d.\ depolarizing noise on the physical qubits.  The effect of depolarizing noise is that each physical qubit is acted upon by the quantum channel
\begin{equation}
 D(\rho_1)=(1-p)\rho_1+\frac{p}{3}\sum_{i=1}^3\sigma^i\rho_1\sigma^i,
\end{equation}
where $0\leq p\leq 1$ is the error probability for the qubit.  

Another important (abelian) group of operators to be considered are the pure errors $\mathcal{E}\subset\mathcal{G}_n$.  This group can be generated by $n-k$ operators $E_i$ that satisfy $E_iS_j=(-1)^{\delta_{ij}}S_jE_i$.  It will be useful later to note that $\mathcal{G}_n$ is generated by all $E_i$, $S_i$, $X_{\alpha}$ and $Z_{\alpha}$.

To detect the presence of errors, we measure the stabilizer generators $S_i$ (each of which has eigenvalues $\pm1$).  The measurement results form the error syndrome $\vec{s}$ where $s_i=\pm 1$ is the outcome of measuring $S_i$.  The syndrome gives us some (but not all) information about any error that may have occurred.  Given an error operator $E$ with syndrome $\vec{s}$ a different error operator $E^{\prime}=LSE$ has the same syndrome for any stabilizer $S\in\mathcal{S}$ and any logical operator $L\in\mathcal{L}$.  
Figuring out which correction operator to apply to correct whatever error may have occured is known as decoding, a computationally challenging problem we  will return to in section \ref{sec:Dec}.

\section{Tensor-network error correcting codes}
\label{sec:TNcodes}
Stabilizer codes can be described using tensors \cite{FHM20}.  We represent operators by strings of integers, e.g., the stabilizer $XYZYX\openone Z=\sigma^{1}\otimes\sigma^{2}\otimes\sigma^{3}\otimes\sigma^{2}\otimes\sigma^{1}\otimes\sigma^{0}\otimes\sigma^{3}$ is represented by the string $(1,2,3,2,1,0,3)$.  We then define the tensors
\begin{equation}\label{eq:STdef}
	T(L)_{(g_1,...,g_n)}=\begin{cases}
                     1 \mathrm{\ \ \  if\ }\sigma^{g_1}\otimes...\otimes\sigma^{g_n}\in L\mathcal{S}\\
                     0 \mathrm{\ \ \  otherwise,}
                    \end{cases}
\end{equation}
where $g_j\in\{0,1,2,3\}$ and $L\mathcal{S}$ is the set of all operators of the form $SL$ with $S\in\mathcal{S}$ and where $L$ is a logical Pauli operator.  In other words, $L\mathcal{S}$ is the coset of $\mathcal{S}$ with respect to the logical operator $L$.  For example, $T(\openone)_{(g_1,...,g_n)}$ is non zero only when $\sigma^{g_1}\otimes...\otimes\sigma^{g_n}$ is a stabilizer, so $T(\openone)$ describes the stabilizer group (except for the overall signs of the Pauli strings, but once these are fixed for the generators, they are fully determined for the whole group).  Similarly, $T(X)_{(g_1,...,g_n)}$ describes all representatives of the logical $X$ operator if there is a single logical qubit.  An interesting special case occurs when there are no logical qubits at all, and $\mathcal{S}$ stabilizes a single state.
As a concrete example, consider the [[7,1,3]] Steane code, with stabilizer generators and logical operators summarised in table \ref{table:Steane}.  
For the Steane code, we denote the code tensor by $T^1(L)_{(g_1,...,g_7)}$, which has $64$ nonzero values for each possible $L\in\{\openone,X,Y,Z\}$.

\begin{table}
\begin{center}
  \begin{tabular}{| c | c c c c c c c | }
    \hline
    Qubit & 1 & 2 & 3 & 4 & 5 & 6 & 7 \\ \hline
    $S_1$ & $X$ & $X$ & $\openone$ & $X$ & $X$ & $\openone$ & $\openone$ \\ \hline
    $S_2$ & $\openone$ & $X$ & $X$ & $X$ & $\openone$ & $\openone$ & $X$ \\ \hline
    $S_3$ & $X$ & $\openone$ & $X$ & $X$ & $\openone$ & $X$ & $\openone$ \\ \hline
    $E_1$ & $\openone$ & $\openone$ & $Z$ & $Z$ & $\openone$ & $\openone$ & $\openone$ \\ \hline
    $E_2$ &  $Z$ & $\openone$ & $\openone$ & $Z$ & $\openone$ & $\openone$ & $\openone$\\ \hline
    $E_3$ &$\openone$ & $Z$ & $\openone$ & $Z$ & $\openone$ & $\openone$ & $\openone$ \\ \hline
    $X_1$ & $X$ & $X$ & $X$ & $X$ & $X$ & $X$ & $X$ \\ \hline
    $Z_1$ & $Z$ & $Z$ & $Z$ & $Z$ & $Z$ & $Z$ & $Z$ \\ 
    \hline
  \end{tabular}
\end{center}
	\caption{Stabilizer generators, pure errors and logical operators for the [[7,1,3]] Steane code.  Note that $S_i$ and $E_i$ for $i\in\{4,5,6\}$ are not displayed as they are the same as $S_i$ and $E_i$ for $i\in\{1,2,3\}$ but with $X$s and $Z$s swapped.  Note that these representations are not unique, e.g., $X_1$ is equivalent to $X_1S_2$ since they act the same way on the codespace.}
\label{table:Steane}
\end{table}

The useful aspect of using these tensors to describe stabilizer codes is that we can combine several tensors together by contracting tensor indices to get new stabilizer codes.
Then if the resulting tensor network can be efficiently contracted, the code can also be efficiently decoded using the method in section \ref{sec:TN_approach}.  
We will apply this decoder in section \ref{sec:dec_holo} to holographic codes, in which case the decoder is provably efficient in the number of physical qubits \cite{FHM20}.  An advantage of using this tensor-network decoder is that we can efficiently decode different logical qubits in parallel, which we prove in section \ref{sec:opt_dec}.

\begin{figure*} [ht!]
\includegraphics[width=\textwidth]{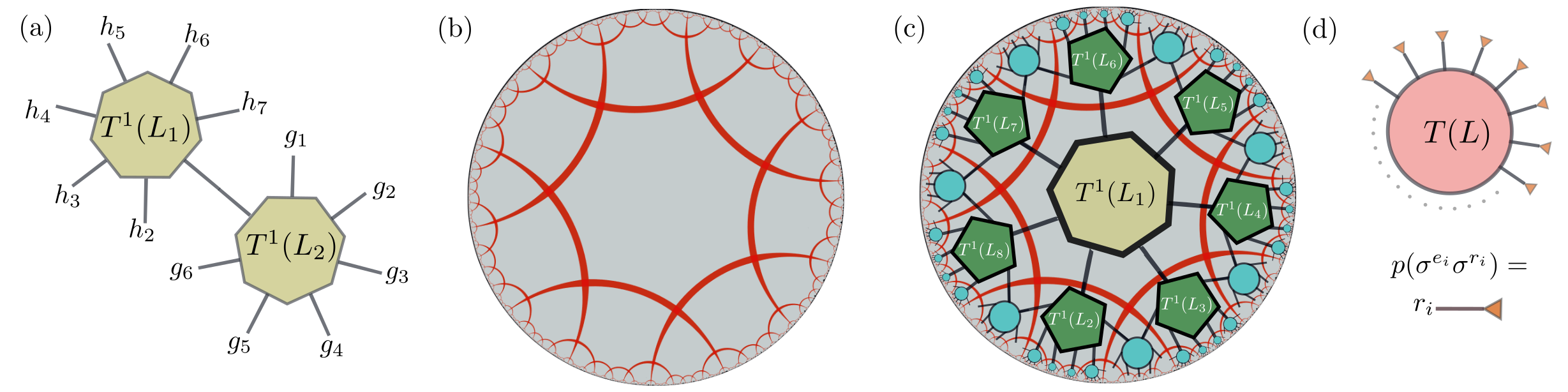}
	\caption{(a) shows a tensor-network code composed of two Steane-code tensors, giving a [[12,2,3]] code.  The contraction of two indices of the Steane code is possible because the Steane code can distinguish single-qubit Pauli errors on any site (because it can correct any single-qubit error).
	The reason this allows us to contract tensor indices to get a new code is discussed around equation (\ref{eq:joining}).
	(b) shows a uniform tiling of hyperbolic space with regular heptagons where four heptagons meet at any vertex.  
	%(Tiling images generated using KaleidoTile \cite{Weeks20}.)
	(c) shows the holographic Steane code.  This is also called the heptagon code because each tensor is naturally associated to a heptagonal tile (though one could conceive of other codes corresponding to a heptagonal tiling not based on the Steane code).
	The code can be constructed by starting from the central tile, which has a Steane-code tensor $T^1(L)$ associated to it.  Then at radius two we add seven further tensors (the same tensors, but now represented by green heptagons), each with one leg contracted to one of the outgoing legs of the central tensor.  
	Finally, we add further $T^1(L)$ tensors (blue discs) at radius three, with one or two outgoing legs contracted with outgoing legs of tensors at radius two.  
	This is a radius-three holographic code.  
	The uncontracted legs at the boundary correspond to physical qubits. 
	The resulting tensor describes a stabilizer code, which follows using equation (\ref{eq:joining}) and noting that each Pauli error localized only on qubits $6$ and $7$ of the Steane code has a unique syndrome.
	(d) shows a tensor-network contraction used for the maximum-likelihood decoder.  The central tensor describes the logical cosets of the code, while the boundary (single-leg) tensors describe the (uncorrelated) noise model on the physical qubits.}
	\label{fig:TN}
\end{figure*}

We can build larger codes using small code tensors as building blocks.  These tensors need not be the same, and not all need to have logical qubits.  
This allows us to iteratively build up very large codes with a guarantee of consistency. A simple example is shown in figure \ref{fig:TN}.  To see how this works, consider two stabilizer-code tensors $T(L)_{(g_1,...,g_{n})}$ and $T^{\prime}(L^{\prime})_{(h_1,...,h_{n^{\prime}})}$ which have $n$ and $n^{\prime}$ physical qubits and $k$ and $k^{\prime}$ logical qubits respectively.
Assume that at least one of these codes can distinguish any Pauli error on a set of qubits, meaning there is a unique syndrome for each Pauli error on those qubits (for simplicity, let us choose qubits $1$ to $l$).  Then we can construct a new tensor describing a new stabilizer code by contracting indices:
 \begin{equation}\label{eq:joining}
 \begin{split}
& T_{\mathrm{new}}(L_{\mathrm{new}}) = \\
	 & \sum_{j_1,...,j_l\in\{0,1,2,3\}}\!\!\!\!\!\!\!\!\!\!\! T(L)_{(j_1,...,j_{l},g_{l+1},...,g_{n})}T^{\prime}(L^{\prime})_{(j_1,...,j_{l},h_{l+1},...,h_{n^{\prime}})},
\end{split}
 \end{equation}
 where $L_{\mathrm{new}}=L\otimes L^{\prime}$ represent logical operators for the new code.  $T_{\mathrm{new}}(L_{\mathrm{new}})$ describes a stabilizer code with $n+n^{\prime} - 2l$ physical qubits and $k+k^{\prime}$ logical qubits.  This is proved in \cite{FHM20}. 

Our main example of tensor-network codes is a specific holographic stabilizer code called the max-rate Steane (or heptagon) holographic code \cite{HMBS18}, which is best understood via figure \ref{fig:TN}.  
We start with a central Steane-code tensor.  Then we contract each outgoing leg of this tensor with another Steane-code tensor, each of which has one ingoing leg contracted with an outgoing leg of the central tensor.  We call this a radius-two code.  To get a radius-three code, we contract tensors with each outgoing leg of the radius-two tensors, but now some radius-three tensors have two legs contracted with two neighbouring radius-two tensors, as shown in figure \ref{fig:TN}.
For this to give a valid stabilizer code, it is enough that the Steane code can distinguish all two-qubit errors on the qubits corresponding to the ingoing legs.  Using this construction, each tensor (and hence each logical qubit) is naturally associated to a tile in a tiling of the hyperbolic plane with heptagons, where four heptagons meet at each vertex.  The physical qubits are on the boundary of this tiling corresponding to the uncontracted legs of the tensor network, as shown in figure \ref{fig:TN}.

\section{Maximum likelihood decoding via tensor networks}
\label{sec:Dec}
Here we will look at maximum likelihood decoding, and we will see in section \ref{sec:opt_dec} that decoding multiple logical qubits can sometimes be done in parallel.  
In section \ref{sec:TN_approach}, we will consider our tensor-network approach to maximum likelihood decoding, comparing it to previous approaches.   
We will see that this tensor-network method does allow us to decode different logical qubits in parallel, which relies on theorem \ref{th:main} in section \ref{sec:opt_dec}.

Provided we know the error model for the physical system, the optimal decoder is the maximum likelihood decoder.  This calculates the error correction operator that is most likely to return to the correct code state given the syndrome.  (Recall that we are not considering the possibility of faulty syndromes.)  
After measuring the stabilizers to get the syndrome $\vec{s}$, we can easily find a pure error $E(\vec{s\,})\in\mathcal{E}$, which is consistent with the syndromes.  Recall, however, that this is not the only possibility:\ any error consistent with syndrome $\vec{s}$ has the form $E(\vec{s}\,)SL$ for some logical operator $L\in\mathcal{L}$ and some stabilizer $S\in\mathcal{S}$ since $\mathcal{S}$, $\mathcal{E}$ and $\mathcal{L}$ generate $\mathcal{G}_n$.  Given an error $E(\vec{s}\,)SL$, then a suitable correction operator would be the inverse operator $LSE(\vec{s}\,)$.  However, suppose that the error that actually occurred was $E(\vec{s}\,)S^{\prime}L^{\prime}$ for some other stabilizer $S^{\prime}$ and logical operator $L^{\prime}$.  Then the net result of our correction will be
\begin{equation}
 [LSE(\vec{s}\,)][E(\vec{s}\,)S^{\prime}L^{\prime}]\ket{\psi}=LL^{\prime}SS^{\prime}\ket{\psi}=LL^{\prime}\ket{\psi},
\end{equation}
where $\ket{\psi}$ is a state in the logical subspace.  If $L\neq L^{\prime}$, then the net result of the error followed by the correction is actually a logical error on the system:\ we have inadvertently applied an \emph{unknown} logical operator.

To minimize the probability of a logical error, the goal of the maximum likelihood decoder is to find out which correction operator is the most likely to correct the error without introducing a logical error.  
Since $E(\vec{s}\,)SL$ acts the same way on the codespace for any $S$, we must sum over the stabilizers to find the probability that some error of the form $E(\vec{s}\,)SL$ occurred, so we want to calculate
\begin{equation}\label{eq:1}
 \chi(L,\vec{s}\,)=\sum_{S\in\mathcal{S}} \mathrm{prob}(E(\vec{s}\,)SL)
\end{equation}
for each logical operator $L\in\mathcal{L}$.  Note that the syndrome $\vec{s}$ is fixed.  Thus, $\mathrm{prob}(E(\vec{s}\,)SL)$ is the probability that the error $E(\vec{s}\,)SL$ acts on the physical qubits.  We should then apply the correction operator $\overline{L}E(\vec{s}\,)$, where $\overline{L}=\argmax_L\chi(L,\vec{s}\,)$.  
Note that the $\chi(L,\vec{s}\,)$ satisfy
\begin{equation}
 \sum_{L\in\mathcal{L}}\sum_{\vec{s}}\chi(L,\vec{s}\,)=1
\end{equation}
since the probabilities must all sum to one.
It is also useful to consider these probabilities conditioned on syndrome $\vec{s}$
This is given by
\begin{equation}\label{eq:zx2}
 \mathrm{prob}(L|\vec{s}\,)=\frac{\chi(L,\vec{s}\,)}{\sum_{L^{\prime}\in\mathcal{L}}\chi(L^{\prime},\vec{s}\,)}
\end{equation}
since
\begin{equation}\label{eq:zx1}
 \mathrm{prob}(\vec{s}\,)=\sum_{L^{\prime}\in\mathcal{L}}\chi(L^{\prime},\vec{s}\,).
\end{equation}

\subsection{Calculating the decoder's success rate}
\label{sec:calc_thresh}
In simulations, since we know the simulated error $E$ that has occurred (unlike in practice), we can check whether the decoder will successfully correct that error.  
In other words, given $E$ we know whether the decoder will or will not introduce a logical error after correction as we can check if the product of the correction operator and the error operator introduces a logical error.  
Thus, we can introduce a success function $\lambda(E)$ with $\lambda(E)=1$ if the decoder succeeds and $\lambda(E)=0$ if it fails.  

The success probability $p_{\mathrm{success}}(p)$ is a function of the single-qubit error probability $p$, and can be estimated in two different ways.  
The first way involves randomly choosing an error based on the physical error probability distribution $\mathrm{prob}(E)$, where $E$ is any possible Pauli error string arising from the noise, i.e., $E=\sigma^{e_1}\otimes...\otimes\sigma^{e_n}$.  We calculate the syndrome $\vec{s}$ and the pure error $E(\vec{s}\,)$.  Then we use the decoder to find the correction operator $LE(\vec{s}\,)$ that maximizes $\mathrm{prob}(L|\vec{s}\,)$.   
We then check if the result is a success or a failure, i.e., we decode successfully if $LE(\vec{s}\,)E\in\mathcal{S}$ in which case $\lambda(E)=1$ (otherwise, $\lambda(E)=0$).  
Monte Carlo sampling over the physical error distribution gives us an approximation to
\begin{equation}\label{eq:Monte1}
 p_{\mathrm{success}}(p)=\sum_{E}\mathrm{prob}(E)\lambda(E),
\end{equation}
where $E$ includes all possible Pauli strings.

But there is a second way to calculate the $p_{\mathrm{success}}(p)$ \cite{BSV14}.  We can use the fact that the decoder itself calculates the probability of successfully decoding given the syndrome $\vec{s}$.  We use
\begin{equation}
\begin{split}
 p_{\mathrm{success}}(p) & =\sum_{E}\mathrm{prob}(E)\lambda(E)\\
 & =\sum_{\vec{s}}\left(\sum_{L\in\mathcal{L}}\sum_{S\in\mathcal{S}}\mathrm{prob}\big(E(\vec{s}\,)SL\big)\lambda\big(E(\vec{s}\,)SL\big)\right),
 \end{split}
\end{equation}
where we used the fact that any Pauli product can be written as $E(\vec{s}\,)SL$ for some $\vec{s}$, $S\in\mathcal{S}$ and $L\in\mathcal{L}$.
Now, the decoder corrects the error $E=E(\vec{s}\,)SL$ successfully only when $L$ is the logical operator that maximizes $\chi(L,\vec{s}\,)$, which we can call $\overline{L}(\vec{s}\,)$.  Then $\lambda(E(\vec{s}\,)SL)=1$ if $L=\overline{L}(\vec{s}\,)$ and is zero otherwise.  So we get
\begin{equation}\label{eq:ab}
\begin{split}
 p_{\mathrm{success}}(p) & =\sum_{\vec{s}}\left(\sum_{S\in\mathcal{S}}\mathrm{prob}\big(E(\vec{s}\,)S\overline{L}(\vec{s}\,)\big)\right)\\
 & =\sum_{\vec{s}}\chi\big(\overline{L}(\vec{s}\,),\vec{s}\,\big)\\
 & =\sum_{\vec{s}}\mathrm{prob}(\vec{s}\,)\mathrm{prob}(\overline{L}(\vec{s}\,)|\vec{s}\,),
 \end{split}
\end{equation}
where the second line follows from equation (\ref{eq:1}), and the last line follows from equations (\ref{eq:zx2}) and (\ref{eq:zx1}).
To use equation (\ref{eq:ab}) to calculate $p_{\mathrm{success}}(p)$, we can do Monte Carlo sampling of syndromes $\vec{s}$ according to $\mathrm{prob}(\vec{s}\,)$ by first sampling errors $E$ according to $\mathrm{prob}(E)$ 
since
\begin{equation}
\begin{split}
 \mathrm{prob}(\vec{s}\,) & =\sum_{L\in\mathcal{L}}\sum_{S\in\mathcal{S}}\mathrm{prob}(E(\vec{s}\,)SL)\\
	& =\sum_{\substack{E=E(\vec{s}\,)SL\\ \forall S \in \mathcal{S}\\ \forall L \in \mathcal{L}}}\mathrm{prob}(E).
 \end{split}
\end{equation}
So, by sampling errors $E$ according to $\mathrm{prob}(E)$, we can either use equation (\ref{eq:ab}) or equation (\ref{eq:Monte1}) to estimate $p_{\mathrm{success}}(p)$.  In both cases, the decoder calculates $\chi(L,\vec{s}\,)$.

Having two different methods gives us a useful consistency check for our estimates for the success probability.  Furthermore, the latter method (using equation (\ref{eq:ab})) has smaller error bars in the holographic-code examples that we will consider in section \ref{sec:dec_holo}. 
In fact, it makes sense that the latter method would be more precise in general because it already has some averaging for a single sample:\ we find the probability of successfully decoding given a syndrome $\vec{s}$, but this includes all errors consistent with that syndrome.  In contrast, a single sample using the first method (i.e., using equation (\ref{eq:Monte1})) only considers a single error operator.

\begin{comment}
Finally, note that what the decoder actually calculates is $\chi(L,\vec{s}\,)$, so it would also be possible to choose syndromes \emph{uniformly at random}, then calculate $\chi(L,\vec{s}\,)$ for each $L$ via the tensor network, find the maximum $\chi(L,\vec{s}\,)$ and use the second line of equation (\ref{eq:ab}) to estimate $p_{\mathrm{success}}$.  \textcolor{blue}{[Is this a good idea?]}
\end{comment}

\subsection{Parallel decoding of logical qubits}
\label{sec:opt_dec}
Suppose we can calculate $\mathrm{prob}(L|\vec{s}\,)$ for each logical operator combination $L$ and a known syndrome $\vec{s}$.  
For $k$ logical qubits, there are $4^k$ such probabilities in the full joint probability distribution to calculate to decode all the logical qubits.  
However, there is an easier way:\ we can decode each logical qubit separately, meaning we have to find $4k$ marginal probabilities (which is straightforward for tensor-network codes as we will see in the following section).  We will show that this is optimal when physical error rates are below threshold (for large enough codes).
All of the following also applies if we only decode a subset of the full number of logical qubits.

Suppose we just want to decode the $j$th logical qubit, then we need to calculate
\begin{equation}
\begin{split}
 \chi_j(L_{j},\vec{s}\,) & =\sum_{L_1,...,L_{j-1},L_{j+1},...,L_k}\chi(L_{1}...L_{k},\vec{s}\,),
 \end{split}
\end{equation}
where we are summing over all logical operators on the right hand side, except for those on logical qubit $j$.  
We use $L_m\in\mathcal{L}_m=\{\openone,X_m,Y_m,Z_m\}$ to denote the set of logical Pauli operators for logical qubit $m$.
Given the $L_{j}$ that maximises $\chi_j(L_{j},\vec{s}\,)$, then the operator $E(\vec{s}\,)L_{j}$ is the  most likely correction to successfully recover logical qubit $j$.  Calculating $\chi_j(L_{j},\vec{s}\,)$ may be difficult in general, but for the tensor networks we will consider it is straightforward.  This is because $\chi_j(L_{j},\vec{s}\,)$ is given by a complete contraction of a tensor network for each $L_{j}$, as we will see in the following section.

We can also introduce marginal conditional probabilities, so for logical qubit $j$ we have
\begin{equation}
 \mathrm{prob}_j(L_j|\vec{s}\,)=\frac{\chi_j(L_j,\vec{s}\,)}{\sum_{L_j^{\prime}\in\mathcal{L}_j}\chi_j(L_j^{\prime},\vec{s}\,)},
\end{equation}
which is the probability, given syndrome $\vec{s}$, that $E(\vec{s}\,)L_j$ will successfully correct logical qubit $j$.  Suppose we independently calculate each $\mathrm{prob}_j(L_j|\vec{s}\,)$ and $\mathrm{prob}_m(L_m|\vec{s}\,)$ for logical qubits $j$ and $m$, and we find that $\overline{L}_j$ and $\overline{L}_m$ maximise these.
This suggests $E(\vec{s}\,)\overline{L}_j\overline{L}_m$ as a correction operator for these two logical qubits. 
Since we have only calculated the \emph{marginal} probability distributions, it is not clear that this is the best choice for correcting both logical qubits simultaneously.
Nevertheless, the following theorems show that calculating the marginals is essentially enough to find the optimal correction operator for any subset of logical qubits.

\begin{theorem}\label{th:main}
Suppose we want to decode $K\leq k$ logical qubits optimally, meaning we want to determine $\argmax_{L}\mathrm{prob}(L|\vec{s}\,)$, where $L=L_1...L_K$ is a logical operator on the $K$ logical qubits.

If the marginal distributions of this subset are sufficiently peaked, i.e., they satisfy
\begin{equation}\label{eq:useful_condition}
 \max_{L_i}\mathrm{prob}_i(L_i|\vec{s}\,)>\frac{K}{K+1},
\end{equation}
where the $i$th marginal distribution is defined by
\begin{equation}
 \mathrm{prob}_i(L_i|\vec{s}\,)=\sum_{L_1,...,L_{i-1},L_{i+1},...L_K}\!\!\!\!\!\!\!\mathrm{prob}(L_1...L_K|\vec{s}\,),
\end{equation}
then the logical operator maximizing $\mathrm{prob}(L|\vec{s}\,)$ is just the product of the logical operators maximizing $\mathrm{prob}_i(L_i|\vec{s}\,)$, i.e.,
\begin{equation}\label{eq:marg-dec}
 \argmax_{L}\mathrm{prob}(L|\vec{s}\,) = \prod_{i=1}^K\argmax_{L_i}\mathrm{prob}_i(L_i|\vec{s}\,).
\end{equation}
This is proved in appendix \ref{app:proof}.
\end{theorem}

To calculate the right hand side of equation (\ref{eq:marg-dec}) we need to calculate $\mathrm{prob}_i(L_i|\vec{s}\,)$ for each logical qubit $i$ and for each of the four $L_i\in\mathcal{L}_i=\{\openone,X_i,Y_i,Z_i\}$.  This amounts to $4K$ calculations (contractions of the tensor network in our case, but this could apply to other methods that calculate $\mathrm{prob}(L|\vec{s}\,)$).  
In contrast, to calculate the joint probability distribution to calculate the left hand side of equation (\ref{eq:marg-dec}) directly requires $4^K$ contractions of the tensor network.
%Note that this is not trivial as the joint probability distribution does not factorize, so we cannot treat each logical qubit as independent.
%Furthermore, each marginal probability distribution $\mathrm{prob}_i(L_i|\vec{s}\,)$ can be calculated in parallel, which means that the time needed for decoding is independent of the number of logical qubits we wish to decode.
Furthermore each of the $4K$ (or $4^K$) computations are independent of each other, and so this problem is embarrassingly parrallel, allowing us to trivially exchange time complexity for space complexity in a classical co-processor.

Theorem \ref{th:main} above relies on the marginal distributions being sufficiently peaked, as described by equation (\ref{eq:useful_condition}).  The following theorem shows that, below the threshold for the logical qubits, the fraction of error instances for which this condition is satisfied tends to one.

\begin{theorem}\label{th:main2}
The fraction of error instances $Q(p)$ when all $K$ logical qubits satisfy equation (\ref{eq:useful_condition}) tends to one as the code gets bigger as long as we are below threshold:
	\begin{equation}
		Q(p) \geq \prod_{i=1}^K\left[p_{\mathrm{success}}^i(p) - K[1-p_{\mathrm{success}}^i(p)]\right],
	\end{equation}
	where $p_{\mathrm{success}}^i(p)$ is the probability of successfully decoding qubit $i$.  Here we have explicitly included the dependence of $Q$ and $p_{\mathrm{success}}^i$ on the error model via a single parameter $p$ (as is the case for i.i.d.\ depolarizing noise), though more general error models will depend on more parameters.
	This is proved in appendix \ref{app:proof}.
\end{theorem}

In the following section, we will discuss the tensor-network approach to calculating $\mathrm{prob}(L|\vec{s}\,)$, and we will see why calculating $\mathrm{prob}_i(L_i|\vec{s}\,)$ for a given $L_i$ is exactly as easy as calculating $\mathrm{prob}(L|\vec{s}\,)$ for a given $L$.

\subsection{Tensor network approach}
\label{sec:TN_approach}
To implement maximum likelihood decoding, we need to calculate $\chi(L,\vec{s}\,)$ for many physical qubits (which is generally difficult \cite{IP13}).  One approach is to use tensor networks to calculate $\chi(L,\vec{s}\,)$ \cite{BSV14,FP14}.  Recall that we wish to calculate
\begin{equation}\label{eq:chi}
 \chi(L,\vec{s}\,)=\sum_{S\in\mathcal{S}} \mathrm{prob}(E(\vec{s}\,)SL).
\end{equation}
The tensor-network approach does not necessarily require us to assume i.i.d.\ noise, and even weakly correlated noise can be considered, but the formulas become somewhat more convoluted. So let us consider the simplest case of i.i.d.\ depolarizing noise to rewrite equation (\ref{eq:chi}) in terms of a tensor network.
As a result, we have
\begin{equation}
 \mathrm{prob}(\sigma^{a_1}\otimes...\otimes\sigma^{a_n})  = \prod_{i=1}^{n}p(\sigma^{a_i}),
\end{equation}
where
\begin{equation}
p(\sigma^{a_i})=\begin{cases}
                                                   1-p\ \ \mathrm{if}\ \ a_i = 0\\
                                                   p/3\ \ \mathrm{otherwise.}
                                                  \end{cases}
\end{equation}
If we write $E(\vec{s}\,)=\sigma^{e_1}\otimes...\otimes\sigma^{e_n}$, then we have
\begin{equation}
\begin{split}\label{eq:2}
	\chi(L,\vec{s}\,) & =\sum_{r_1,...,r_n\in\{0,1,2,3\}}T(L)_{(r_1...r_n)}\prod_{i=1}^{n} p(\sigma^{e_i}\sigma^{r_i}),
 \end{split}
\end{equation}
where $T(L)_{(r_1...r_n)}$ is the stabilizer-code tensor, so $T(L)_{(r_1...r_N)}=1$ if $\sigma^{r_1}\otimes...\otimes\sigma^{r_N}\in L\mathcal{S}$ and is zero otherwise.  
This tensor network is illustrated in figure \ref{fig:TN}, where $p(\sigma^{e_i}\sigma^{r_i})$ is a sinlge-leg tensor corresponding to physical qubit $i$, which has leg index $r_i$.  Note that $e_i$ is actually fixed because it comes from the pure error $E(\vec{s}\,)$, which is fixed by $\vec{s}$.

For big codes $T(L)_{r_1...r_n}$ can be a very complex tensor, so the contraction would be generally intractible for large $n$.  For the surface code, with one logical qubit, $T(L)$ was decomposed into small tensors in \cite{BSV14}.  In \cite{FP14}, the encoding unitary circuit for some specific codes was chosen to have a structure amenable to contraction, and logical and syndrome qubits were decoded sequentially.  In \cite{CF18}, the approach was to recast decoding in terms of tensor-network calculations of partition functions, which also corresponds to maximum likelihood decoding.

\begin{figure*} [ht!]
	\includegraphics[width=\columnwidth]{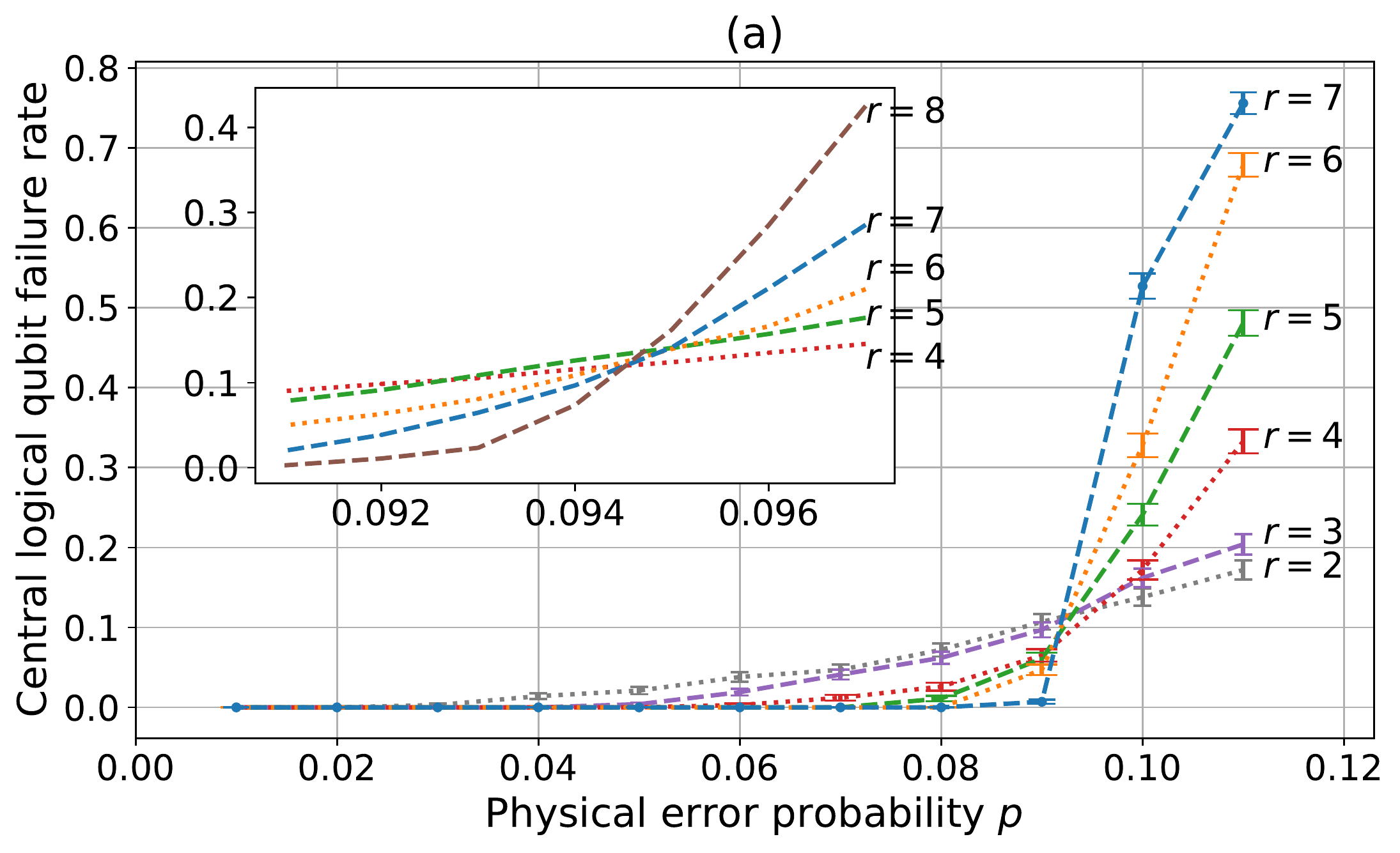}
\includegraphics[width=\columnwidth]{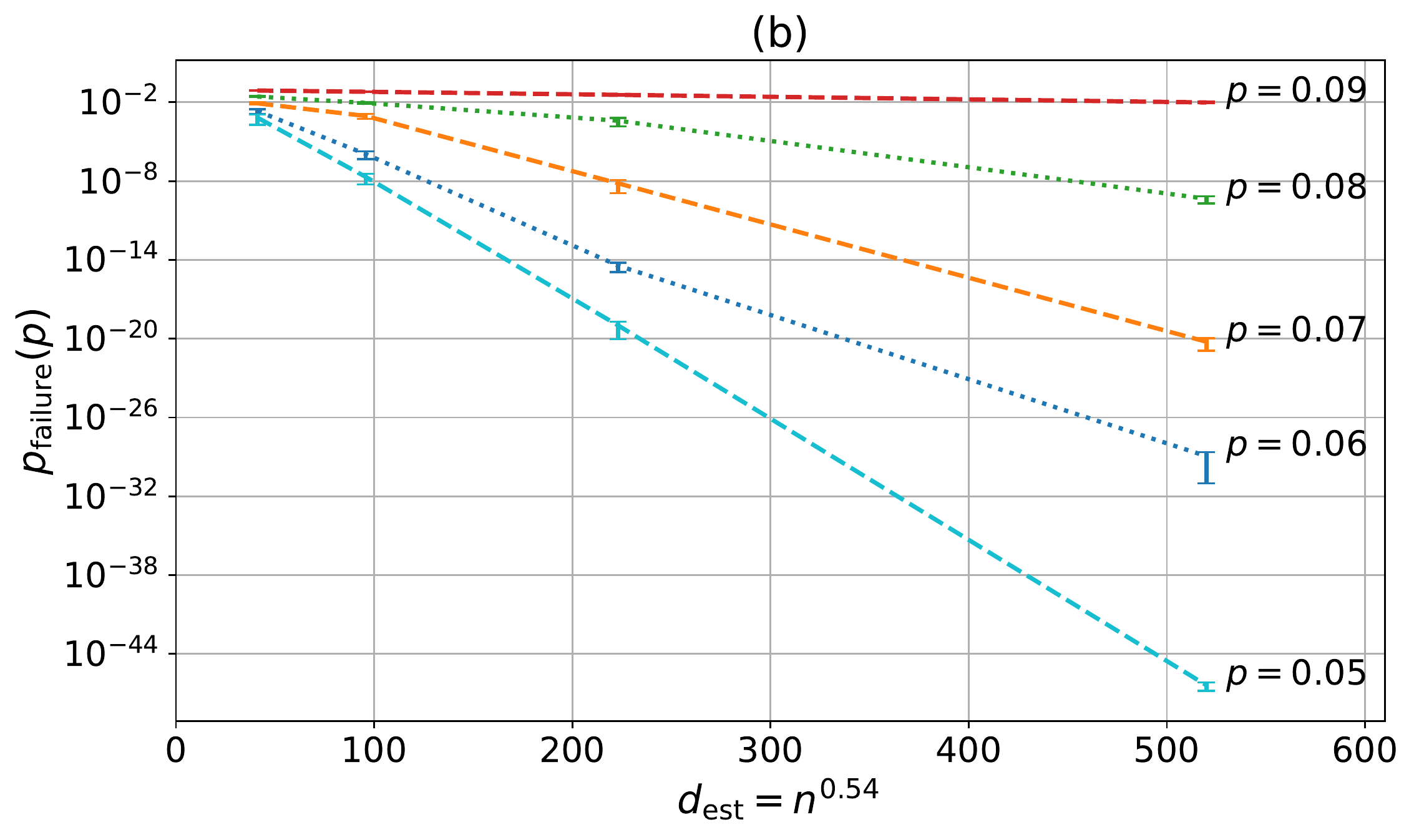}
\includegraphics[width=\columnwidth]{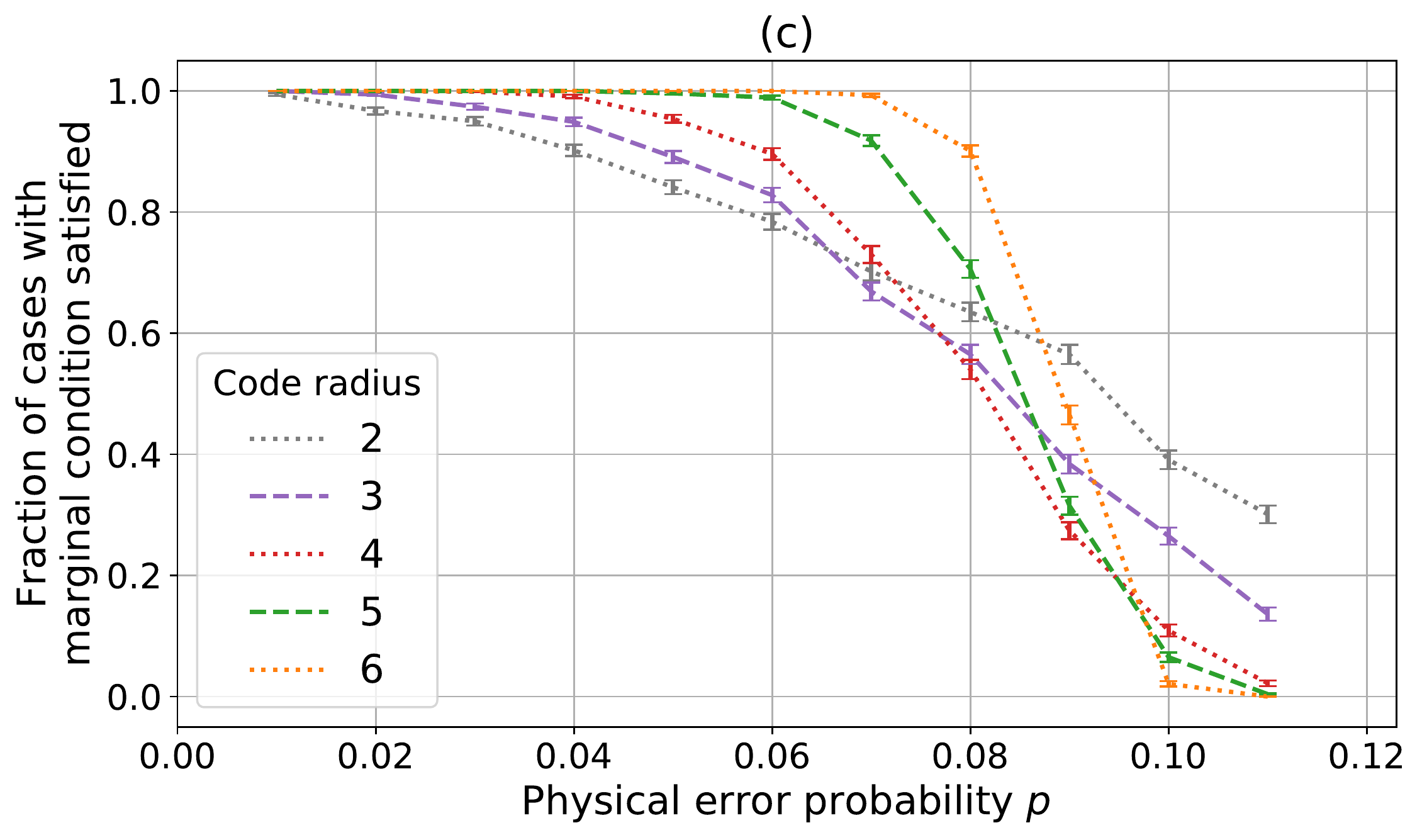}
	\includegraphics[width=\columnwidth]{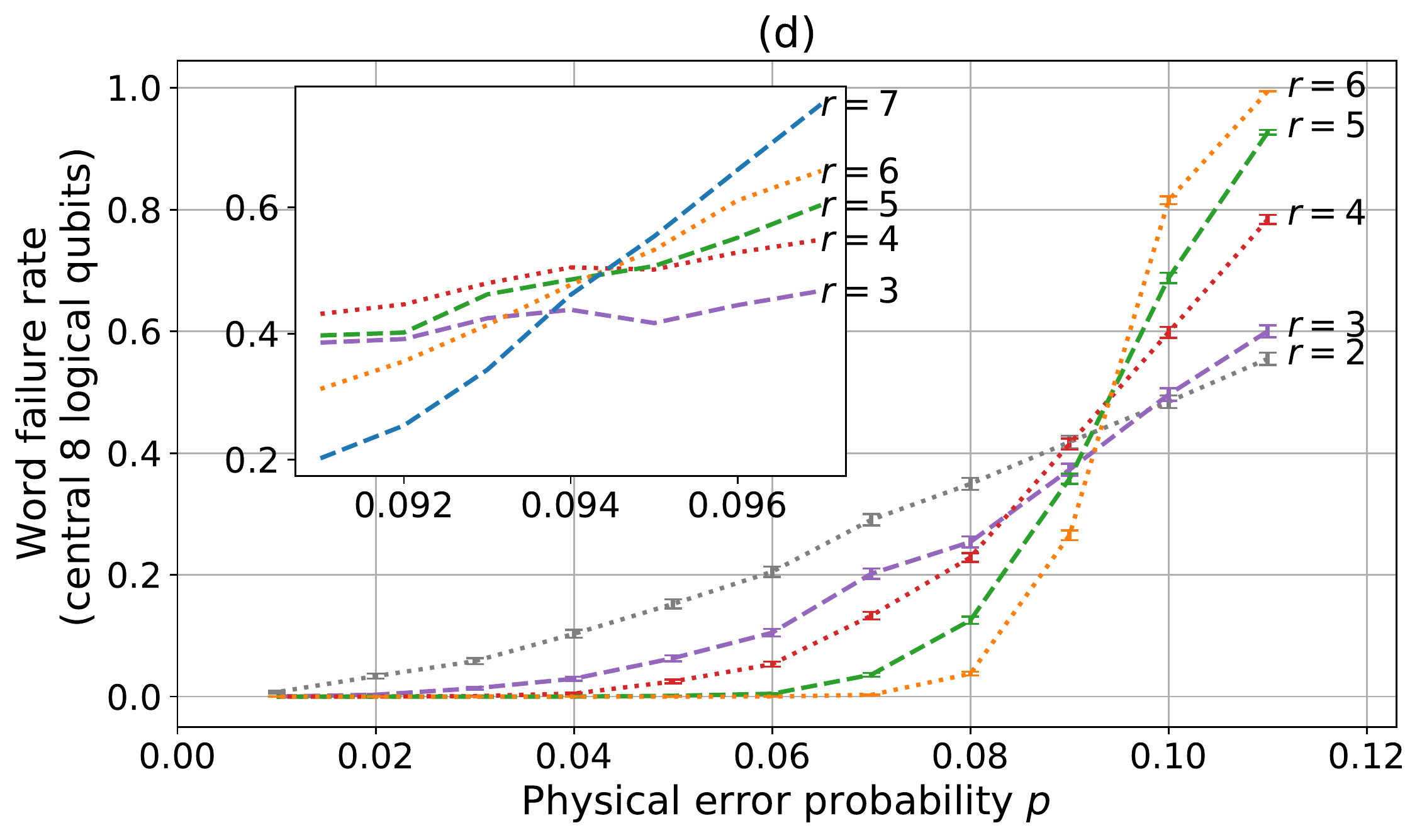}
	\caption{
All points in the above plots correspond to $1000$ samples and error bars are standard errors.
		(a) shows Monte Carlo estimates for the probability of a logical error for the central logical qubit versus the single-qubit error probability for different code radii.  The estimates of $p_{\mathrm{failure}}= 1-p_{\mathrm{success}}$ used the final line of equation (\ref{eq:ab}).  
		The error bars are generally smaller than the error bars when $p_{\mathrm{failure}}$ is estimated using equation (\ref{eq:Monte1}), which is shown in appendix \ref{app:plots} in figure \ref{fig:121}.
	(b) shows $p_{\mathrm{failure}}(p)$ versus $d_{\mathrm{est}}=n^{0.54}$, where the latter is the estimate from \cite{HC20} for the code distance.  Each line corresponds to a different value of the depolarizing noise strength $p$.   This plot is consistent with exponential decay of the logical error rate with the code distance. 
			(c) shows the fraction of cases (instances of the randomly chosen error $E$) for which all eight logical qubits satisfy the necessary condition $\max_{L_i}\mathrm{prob}_i(L_i|\vec{s}\,)>K/(K+1)=8/9$, which guarantees that the marginal decoder is optimal.  (The marginal decoder can still be correct even when this condition is not satisfied.)
		(d) shows Monte Carlo estimates for the word failure probability of the central eight logical qubits.
	}
	\label{fig:6}
\end{figure*}

In our case, in contrast to previous approaches \cite{BSV14,FP14,CF18}, we will decode independent logical qubits in parallel.  To see how this is possible, suppose that the full code tensor $T(L)$ is a product of two smaller tensors, $T_i(L_{i})_{(g_1,...,g_{n_i})}$.  So we have
\begin{equation}
\begin{split}
 & T(L_{1}L_{2})_{g_2,...,g_{n_1},h_2,...,h_{n_2}}\\
 & = \sum_{l\in\{0,1,2,3\}}T_1(L_{1})_{l,g_2,...,g_{n_1}} T_2(L_{2})_{l,h_2,...,h_{n_2}}.
 \end{split}
\end{equation}
To calculate $\chi(L,\vec{s}\,)$ for each logical operator $L=L_1L_2$, we need to contract $T(L_{1}L_{2})$ as in equation (\ref{eq:2}).  But if we only want to decode logical qubit $1$, for example, we need only calculate $\chi_1(L_1,\vec{s}\,)=\sum_{L2}\chi(L_1L_2,\vec{s}\,)$.
\begin{comment}
So we need only calculate
\begin{equation}
\begin{split}
 & \chi(L,\vec{s}\,)\\ & =\sum_{r_1,...,r_n\in\{0,1,2,3\}}\sum_{L_2}T(L)_{r_1...r_{n_1+n_2-2}}\prod_{i=1}^{n} p(\sigma^{e_i}\sigma^{r_i}),
 \end{split}
\end{equation}
\end{comment}
This means we do the same contraction as in equation (\ref{eq:2}), but now with
\begin{equation}
\begin{split}
\sum_{l\in\{0,1,2,3\}}T_1(L_{1})_{l,g_2,...,g_{n_1}} Q_{l,h_2,...,h_{n_2}},
 \end{split}
\end{equation}
where
\begin{equation}
\begin{split}
Q_{h_1,...,h_{n_2}} = \sum_{L_2}T_2(L_{2})_{h_1,...,h_{n_2}}.
 \end{split}
\end{equation}
This means that the tensor network we need to contract just has the tensor $T_2$ replaced by the tensor $Q$, so the geometry of the tensor network is unchanged.  
If the geometry of the tensor network before making this replacement allowed efficient contraction, then after this replacement the tensor network can still be contracted efficiently.  Therefore, $\chi_1(L_1,\vec{s}\,)$ is as easy to calculate as $\chi(L_1L_2,\vec{s}\,)$.  This argument applies to arbitrary tensor-network codes to allow us to calculate $\chi_i(L_i,\vec{s}\,)$ for the $i$th logical qubit.

\section{Decoding holographic codes}
\label{sec:dec_holo}
To decode holographic codes, we need to contract the tensor network describing the full code tensor.  Using the contraction scheme of \cite{FHM20}, this can be done efficiently in the number of physical qubits, allowing us to do Monte Carlo sampling to estimate the probability of successfully decoding for different code radii. 
The number of physical qubits grows quickly with the code radius, and for the largest code we consider (with radius eight), the number of physical qubits is over half a million.  (One method to calculate the number of physical qubits is detailed in the appendix of \cite{FYH15}.)

For the central logical qubit, the results are shown in figure \ref{fig:6} (a).  From this, we would expect the threshold for the central logical qubit $p_{\mathrm{th}}$ to be somewhere around $9.4\%$, which is confirmed in appendix \ref{app:plots}.
We also see in figure \ref{fig:6} (b) that the logical failure probability decreases exponentially below threshold, where we used the estimate for the code distance of $n^{0.54}$ from \cite{HC20}.

As explained in section \ref{sec:opt_dec}, we can use the tensor-network decoder to decode multiple logical qubits in parallel.  We considered the central eight logical qubits, i.e., the central logical qubit at radius $r=1$ and the seven adjacent logical qubits at $r=2$ (as in figure \ref{fig:TN}).  

In theorem \ref{th:main} we saw that the fraction of error instances $Q(p)$ for which parallel decoding of individual logical qubits is optimal tends to one as the code gets bigger below threshold.  In figure \ref{fig:6} (c) we plot $Q(p)$ for the central eight logical qubits for different code sizes and different values of $p$ the single-qubit error rate.  Notice that below threshold $Q(p)$ increases for larger codes, as expected from theorem \ref{th:main}.  
It is important to bear in mind that this result is useful for a fixed number $K$ of logical qubits.  Decoding all the logical qubits of the code will give poor results for, e.g., logical qubits at the boundary.

We expect that any individual logical qubit that is a fixed distance from the centre of the holographic code will have the \emph{same} threshold as the central logical qubit as the code grows due to the symmetry of the code (since the tensors are the same on every tile).  
We can also check this directly:\ figure \ref{fig:6} (d) indeed shows that the word failure rate of the central eight logical qubits has a threshold around $9.4\%$.  Note that we are considering logical qubits at a fixed distance from the code centre.  
In contrast, logical qubits at the boundary will always be highly susceptible to errors and will not have a threshold.

This calculation of the word error probability was simplified because the tensor-network decoder lets us easily calculate the word failure rate for any $K$ logical qubits.  
Once we have found the logical operators $\overline{L}_1,...,\overline{L}_K$ that maximize each individual $\mathrm{prob}_i(L_i|\vec{s}\,)$, we need only perform one extra contraction of the tensor network to calculate $\mathrm{prob}(\overline{L}_1...\overline{L}_K|\vec{s}\,)$.  

As logical qubits close to the boundary are poorly protected from errors, it is natural to ask whether holographic codes can be useful and still have finite rates.  We can answer this question in the affirmative for homogeneous codes, such as the Steane holographic code here.
Consider a logical qubit a distance $\ell$ from the boundary of the code (e.g., in figure \ref{fig:TN} (c), the green qubits are distance $2$ from the boundary).  
We expect that (below threshold) these logical qubits will be at least as well protected from errors as a central logical qubit in a radius $\ell$ code, meaning that the logical failure rates will be less than that of a central logical qubit in a radius $\ell$ code.
There are two reasons to believe this.
The first is that a logical qubit a distance $\ell$ from the boundary in a radius $R$ code is equivalent to starting with a central logical qubit in a radius $\ell$ code and then (asymmetrically) adding more tensors to increase the code size to radius $R$ in such a way that the logical qubit of interest remains a distance $\ell$ from the boundary.
The additional encoding can only increase the distance of this logical qubit's logical operators, so we expect it should be better protected from errors.
The second reason is simply that our numerics corroborate this:\ all logical qubits at radius two in a radius $R$ code had lower logical failure probabilities than a central logical qubit (i.e., at radius one) in a radius $R-1$ code.  This was true for all values of $p\leq 0.8$, but for $p=0.9$ the central logical in a radius $R-1$ code had lower logical failure probability, but this discrepancy went to zero as $R$ was increased (we considered $R\in\{3,4,5,6\}$).

Thus, given a depolarizing error probablity $p$ and a desired minimum logical failure probability $p_f$, we can find an $\ell$ such that all logical qubits a distance of more than $\ell$ from the code boundary have logical failure rate less than $p_f$.  Furthermore, the rate for this code, only including the well-protected logical qubits, is
\begin{equation}
	r(\ell) = \frac{r(0)}{\lambda^{\ell}},
\end{equation}
where $r(0)$ is the code rate including all logical qubits ($r(0)= 1/\sqrt{21}$ for the holographic Steane code \cite{HMBS18}), and $\lambda$ determines how quickly the number of tensors grows with the code radius, i.e., for the holographic Steane code $\lambda \simeq 4.8$.  To derive this formula, we use that the number of physical qubits and the number of logical qubits in a radius $R$ code are both proportional to $\lambda^R$, albeit with different constants of proportionality (see the derivation of rates in appendix C of \cite{FYH15}).

It is worth mentioning, however, that a similar argument can be made for zero-rate codes.  Consider a concatenated code with depolarizing noise strength $p$ and desired logical failure probability $p_f$.  
We can always find a concatenation depth $D$ such that the logical failure probability is below $p_f$ (as long as $p$ is below the code threshold).  Then we can encode any number of logical qubits by concatenating each independently to depth $D$.
For concatenated Steane codes, the rate would then be $r(D)= 1/7^D$.  It is not clear which is superior, concatenated or holographic codes, from this simple argument because $\ell$ and $D$ will depend differently on $p_f$, and for holographic codes, logical qubits closer to the centre are better protected from errors than others, whereas all of the concatenated logical qubits are equally well protected. 

\section{Conclusions}
We showed that the tensor-network decoder for tensor-network codes has the property that individual logical qubits can easily be decoded in parallel.  And we saw that this was close to optimal below threshold.  We applied this parallel decoder to the Steane holographic code to verify this for the central eight logical qubits.  We also calculated the threshold of the logical qubits of the Steane holographic code to be $9.4\%$ under depolarizing noise in contrast to a previous estimate of $7\%$ using a different decoder \cite{HC20}.

A feature of the parallel decoding scheme for holographic codes is that the time needed for decoding is independent of the number of logical qubits we wish to decode.
The overall complexity is $O(k\times\mathrm{poly}(n))$ since the tensor-network decoder applied to holographic codes to decode a single logical qubit has complexity $\mathrm{poly}(n)$ \cite{FHM20}.

\begin{comment}
There are many questions to be answered both regarding holographic codes and tensor-network stabilizer codes more generally.  For example, what tensor-network codes can we construct by contracting code tensors together maybe in three dimensions or even in cases where there is no natural geometry?  What does the set of stabilizer codes look like when factored into products of stabilizer codes?  In other words, when can we decompose a stabilizer code on $n$ qubits into two stabilizer codes with some number of tensor legs contracted?

For holographic codes, how does the decoder perform if we do include some type of bond-dimension truncation?  How do holographic codes perform more generally in a fault-tolerant setting?  (An advantage of holographic code states is that they can be constructed via the cluster-state method \cite{MH20}.)  Finally, how do holographic codes perform in the presence of correlated noise?  If the error channel could be represented by a tensor network with a low enough bond dimension (corresponding to local correlations), then we could contract the tensor network.  This setting was studied for the surface code in \cite{CF18}.
\end{comment}

\acknowledgments
The authors would like to thank Aidan Strathearn and Yoni Nazarathy for useful discussions.
This work was supported by the Australian Research Council Centres of Excellence for Engineered Quantum Systems (EQUS, CE170100009) and the Asian Office of Aerospace Research and Development (AOARD) grant FA2386-18-14027.
Numerical simulations were performed on The University of Queensland's School of Mathematics and Physics Core Computing Facility ``getafix'' (with thanks to Dr. L. Elliott and I. Mortimer for computing support).

\bibliographystyle{unsrt}
%\bibliography{../References}

\appendix

\section{Proof of theorem \ref{th:main} and theorem \ref{th:main2}}\label{app:proof}
To prove theorem \ref{th:main}, our goal is to show that, gven a joint probability distribution $p_{i_1,...,i_K}$, if the marginals satisfy
 \begin{equation}
  \max_{i} p^{\alpha}_{i}> \frac{K}{K+1}
 \end{equation}
for each marginal labelled by $\alpha\in\{1,...,K\}$, then
 \begin{equation}
  \argmax_{i_1,...,i_K} p_{i_1,...,i_K} = (\argmax_{i_{1}} p^1_{i_1},...,\argmax_{i_{K}} p^K_{i_K}).
 \end{equation}
 In other words, to find where the maximum of $p_{i_1,...,i_K}$ is, it is sufficient to find the maxima of the marginals $p^{\alpha}_{i}$.

\begin{proof}
Marginals are defined by
\begin{equation}
 p^{\alpha}_i = \sum_{j_1,...,j_{\alpha-1},j_{\alpha+1},...,j_K}p_{j_1,...,i_{\alpha},...,j_K}.
\end{equation}
	For simplicity, let us suppose that each marginal has its maximum value at index $0$, which implies that $p^{\alpha}_0>K/(K+1)$.

We can prove lemma 1 by contradiction:\ suppose the maximum of $p_{j_1,...,j_K}$ occurs at an index not equal to $(0,...,0)$.  
Now define
\begin{equation}
	\begin{split}
		x_0 & = p_{0,...,0}\\
		x_1 & = \sum_{j_1 \geq 1}\sum_{j_2,...,j_K}p_{j_1,...,j_K}\\
		x_2 & = \sum_{j_2 \geq 1}\sum_{j_3,...,j_K}p_{0,j_2,...,j_K}\\
		& ...\\
		x_K & = \sum_{j_K \geq 1}p_{0,...,0,j_K}.
	\end{split}
\end{equation}
	Note that $\sum_i x_i = 1$.  Furthermore, we have
	\begin{equation}
		\begin{split}
			\sum_{l\neq m}x_l & \geq \!\!\!\! \sum_{j_1,...,j_{m-1},j_{m+1},...,j_K} \!\!\!\! p_{j_1,...,j_{m-1},0,j_{m+1},...,j_K}\\
			& = p^m_0 >\frac{K}{K+1}
		\end{split}
\end{equation}
	for each $m\neq 0$.  Let the largest $x_m$ have index $m^*$, which we know cannot be zero since we assumed that $p_{0,...,0}$ was not the largest element of the probability distribution.  Then we have
\begin{equation}
 \sum_{l\neq m_*}x_l=1-x_{m^*}>\frac{K}{K+1},
\end{equation}
which means that $x_{m^*}<1/(K+1)$, but we also have
\begin{equation}
 kx_{m^*}\geq\sum_{l\neq m}x_l>\frac{K}{K+1},
\end{equation}
which follows because $x_{m^*}$ is the largest element.  This implies $x_{m^*}>1/(K+1)$, which is a contradiction, as promised.

The lower bound on the maxima of the marginals $p^{\alpha}_i$ in the lemma ($K/(K+1)$) is optimal, in the sense that any other lower bound does not guarantee that the marginals allow us to find the global maximum.  This can be seen from an example with $p_{0,..,0}=1/(K+1)-\epsilon$ and $p_{1,0,...,0}=...=p_{0,...,0,1} =1/(K+1)+\epsilon/K$ for some small $\epsilon>0$.  Then the marginals satisfy $p^{\alpha}_0=K/(K+1)-\epsilon/K$ but $p_{0,..,0}$ is not the biggest value of the full distribution.
\end{proof}

To prove theorem \ref{th:main2}, we want to lower bound the fraction of error instances where each marginal distribution satisfies $\max\mathrm{prob}(L_i|\vec{s}\,)>K/(K+1)$.

Consider logical qubit $i$, and let $p_{\mathrm{success}}^i$ be the probability of successfully correcting qubit $i$.
We know that
\begin{equation}
 p_{\mathrm{success}}^i = \sum_{\vec{s}}\mathrm{prob}(\vec{s}\,)\mathrm{prob}(\overline{L}_i|\vec{s}\,),
\end{equation}
where $\overline{L}_i=\argmax_{L_i}\mathrm{prob}(L_i|\vec{s}\,)$ for a given syndrome $\vec{s}$.  Let $q_i$ be the probability that $\vec{s}$ corresponds to a case with $\mathrm{prob}(\overline{L}_i|\vec{s}\,)>K/(K+1)$, i.e.,
\begin{equation}
	q_i = \sum_{\substack{\vec{s}\\ \mathrm{prob}(\overline{L}_i|\vec{s}\,)>K/(K+1)}}\mathrm{prob}(\vec{s}\,).
\end{equation}
This allows us to upper bound $p_{\mathrm{success}}^i$ as follows. 
\begin{equation}
\begin{split}
 p_{\mathrm{success}}^i & \leq q_i + (1-q_i)\frac{K}{K+1}.
 \end{split}
\end{equation}
Rearranging, we get
\begin{equation}
 q_i \geq p_{\mathrm{success}}^i - K(1-p_{\mathrm{success}}^i).
\end{equation}
Then a lower bound on the probability that all logical qubits satisfy $\max\mathrm{prob}(L_i|\vec{s}\,)>K/(K+1)$ is
\begin{equation}\label{eq:marg_condition}
	Q \geq \prod_{i=1}^K q_i = \prod_{i=1}^K\left[p^i_{\mathrm{success}} - K(1-p^i_{\mathrm{success}})\right].
\end{equation}

Below threshold $p_{\mathrm{success}}^i$ becomes larger for larger codes, and therefore so does $q$, the fraction of time that logical qubit $i$ satisfies the criterion for the parallel decoder to be optimal.  Suppose we are considering a \emph{fixed} number of logical qubits for bigger and bigger codes.  It follows from equation (\ref{eq:marg_condition}) that, as long as we are below the threshold for the individual logical qubits, the fraction of cases when decoding the logical qubits individually is optimal $Q$ will tend to one.  (Note that decoding the individual qubits can also still give the right answer when $\max\mathrm{prob}(L_i|\vec{s}\,)>K/(K+1)$ is not satisfied.)

\section{Threshold}
\label{app:thresh}
\begin{figure}[ht!]
\includegraphics[width=\columnwidth]{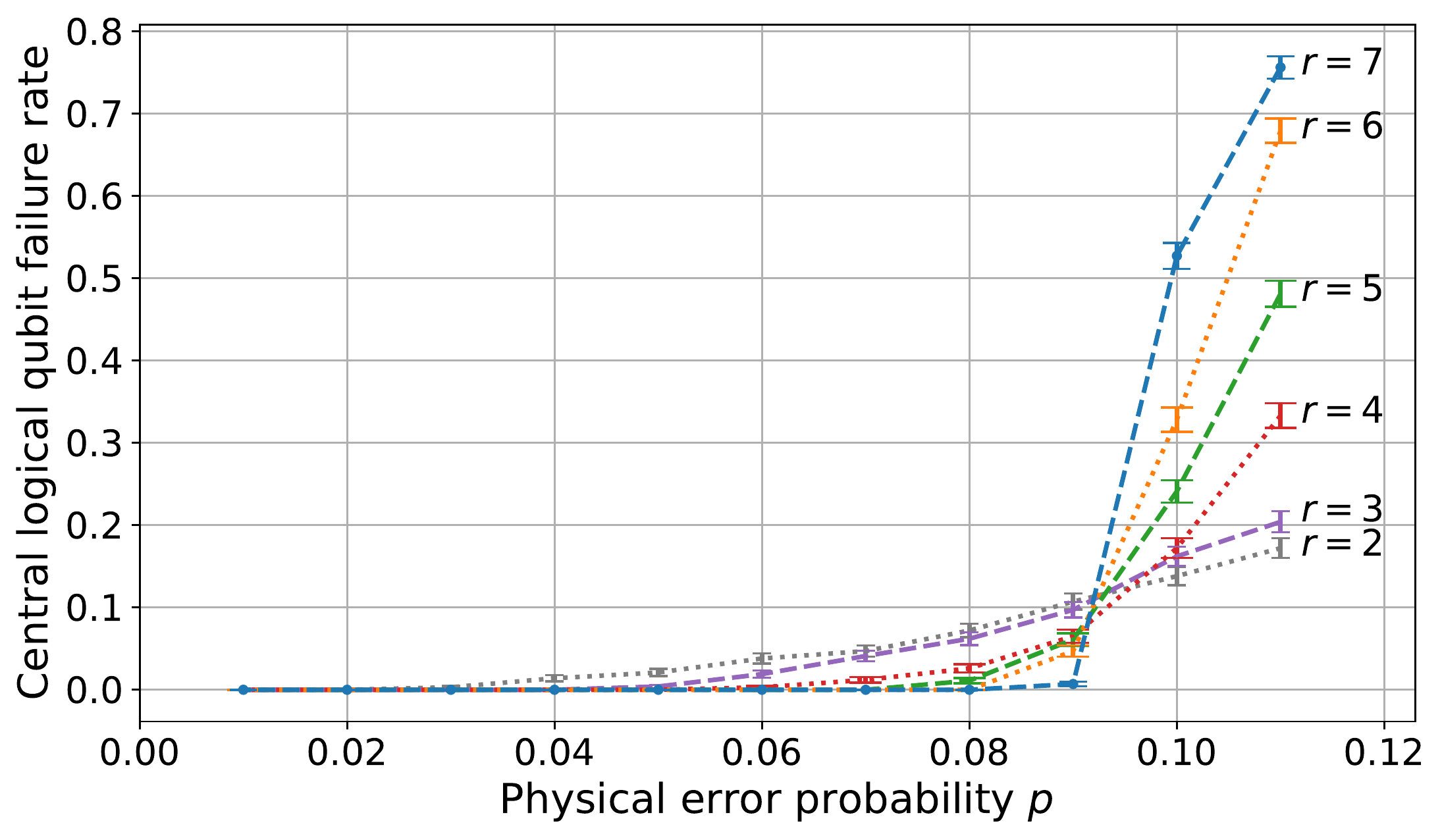}
	\caption{This shows estimates for the probability of a logical error $p_{\mathrm{failure}}=1-p_{\mathrm{success}}$ when decoding the central logical qubit plotted against the single-qubit error probability for different code radii.  For each point we have taken $1000$ samples.  This was obtained by using Monte Carlo sampling to estimate $p_{\mathrm{failure}}$ via equation (\ref{eq:Monte1}).  Error bars correspond to standard errors, but are generally larger than in figure \ref{fig:6} (a).}
\label{fig:121}
\end{figure}

In section \ref{sec:calc_thresh}, we explained that there are two methods to estimate the probability of successfully decoding, the first of which used Monte Carlo sampling of errors in conjunction with equation (\ref{eq:Monte1}).  
Figure \ref{fig:121} shows the result of using this method for the holographic Steane code when decoding only the central logical qubit.  In comparison with the second method (using Monte Carlo sampling of errors together with equation (\ref{eq:ab})) shown in figure \ref{fig:6}, the error bars are somewhat larger.
\label{app:plots}

\begin{figure}[ht!]
\includegraphics[width=\columnwidth]{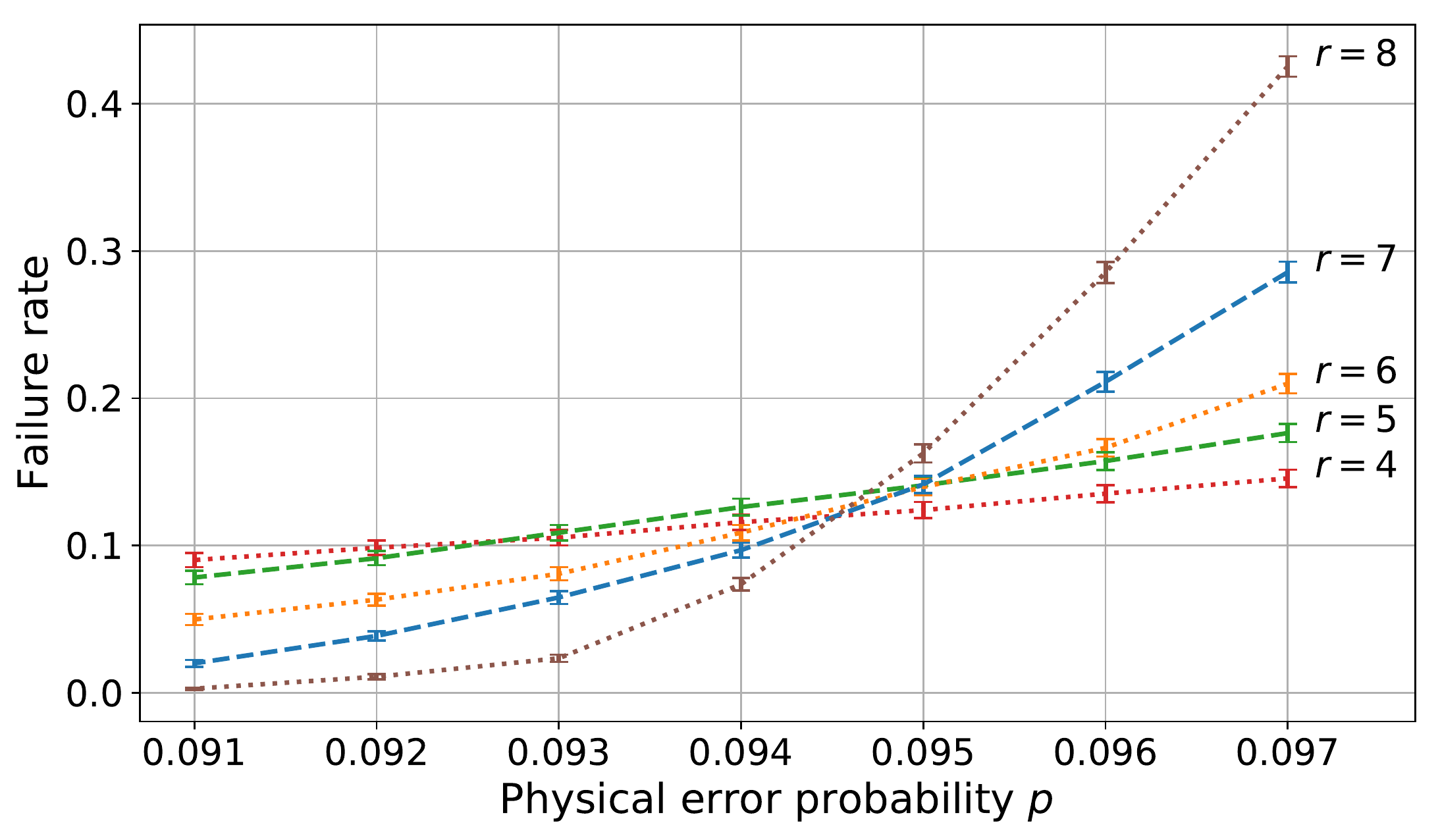}
\caption{
	Monte Carlo estimates of the probability of failing to correct an error for the central logical qubit of the holographic Steane code as a funciton of the single-qubit depolarizing probability $p$.  Each point corresponds to $1000$ samples, and error bars are standard errors.
	}\label{fig:323}
\end{figure}
Figure \ref{fig:323} shows the failure probability for the central logical qubit close to the threshold for varius different code radii.  
Upon inspection, it the threshold looks to be just below $9.5\%$ .  To verify this estimate for the threshold, we employ a scaling hypothesis, following \cite{WHP03}, which looked at the surface code.  We use that, at the threshold $p_{\mathrm{th}}$, the success probability will be scale invariant, meaning it does not depend on the code size.  For any larger error $p>p_{\mathrm{th}}$, a bigger code will perform worse, whereas for a smaller error $p<p_{\mathrm{th}}$, bigger codes will perform better.

\begin{figure}[ht!]
\includegraphics[width=\columnwidth]{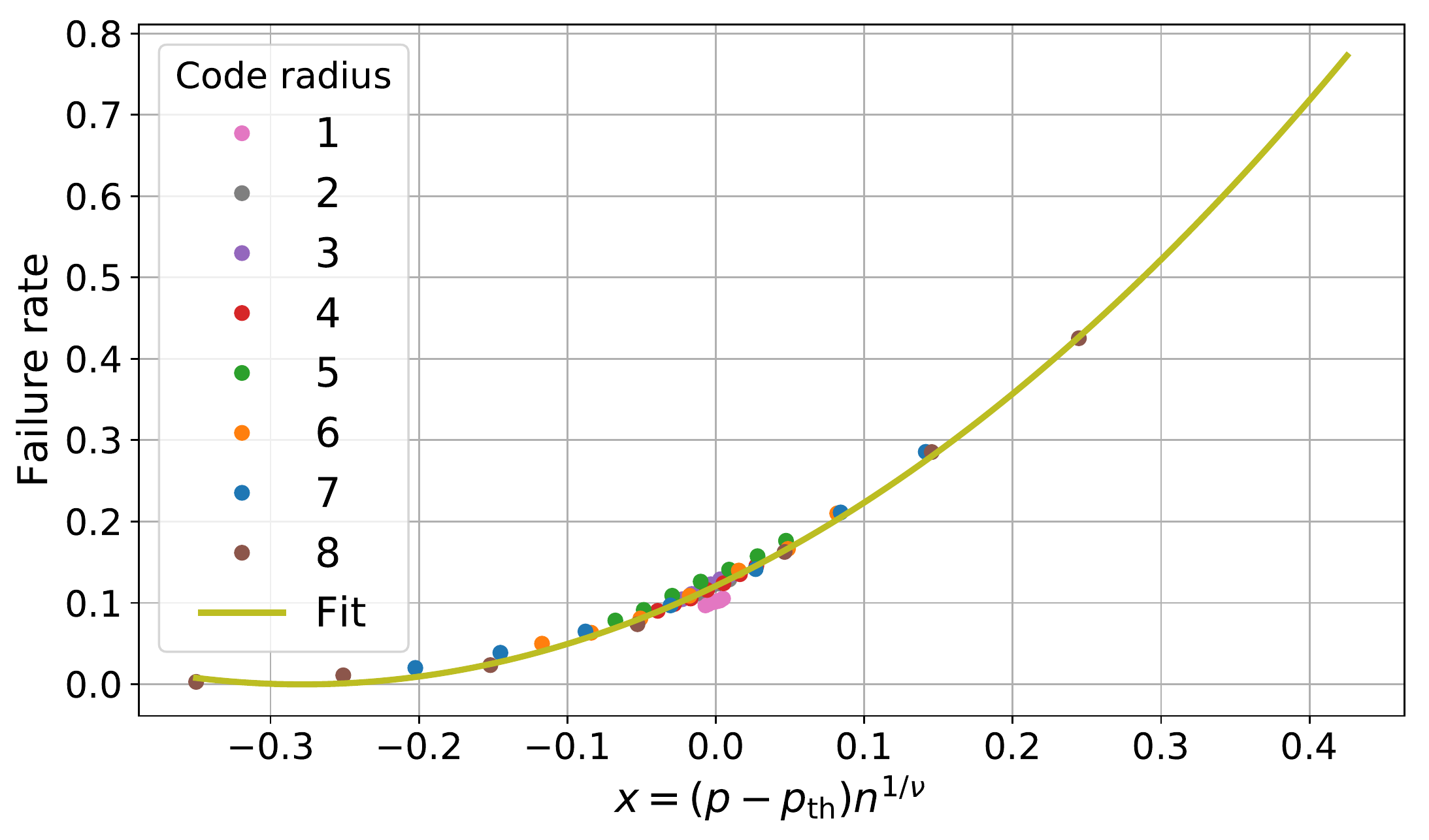}
\caption{
	Failure  rate  for  the  Steane  holographic code's central logical qubit now plotted against the rescaled error probability $x= (p - p_{\mathrm{th}})n^{1/\nu}$ for code radii from $1$ to $8$.  
	We fit a quadratic (using least squares fitting) to the radius-eight data to estimate $f(x)$ because the radius-eight code is closest to the large-system limit.  This allowed us to find values of $p_{\mathrm{th}}$ and $\nu$ such that the data for all radii are as close as possible to the curve.  The resulting values are $\nu= 2.96$ and $p_{\mathrm{th}}= 0.09447(5)$.
	}\label{fig:454}
\end{figure}

Close to the critical point, we expect there will be an error correlation length $\xi$.  We assume that this obeys $\xi \sim |p-p_{\mathrm{th}}|^{-\nu}$, where $\nu$ is critical exponent.  
At the threshold, because of scale invariance, $\xi$ will grow with the size of the system system $n$, so we should consider $\xi/n$.  
In figure \ref{fig:454}, we plot the failure probability now as a function of $x=(p-p_{\mathrm{th}})n^{1/\nu}$.
The assumption we make (following \cite{WHP03}) is that there exists a universal function $f(x)$ describing the failure probability for large codes.  
$f(x)$ should satisfy $p_{\mathrm{failure}}\sim 3/4f(x)$ with the property that $f(x)\rightarrow 1$ as $x\rightarrow \infty$ and $f(x)\rightarrow 0$ as $x\rightarrow -\infty$.

To estimate $p_{\mathrm{threshold}}$, we fit a polynomial function to the data for the largest (radius eight) code to get the best approximation for $f(x)$.  Next we calculated the values of $\nu$ and $p_{\mathrm{th}}$ that allow $f(x)$ to best fit the remaining the data (corresponding to the other code radii).  
The resulting value of $p_{\mathrm{th}}$ was $9.447(5)\%$ (with $\nu=2.96$).  The optimization was done using the LsqFit package in Julia.

The same method can be applied to each of the outer seven logical qubits at radius two to get their thresholds.  In that case, we get a minimum threshold of $9.42(2)\%$ and a maximum of $9.46(1)\%$, so these all agree well considering the error estimates.  Furthermore, as expected, the word error threshold for the eight logical qubits (one at radius one and seven at radius two) agrees, giving a value of $9.42(1)\%$.

\end{document}